\documentclass[aps,prd,showpacs,showkeys,amsfonts,amsmath,reprint,floatfix]{revtex4-1}  

\usepackage[utf8]{inputenc}
\usepackage[T1]{fontenc}
\usepackage{hyperref}
\usepackage{bm}
\usepackage{graphicx}
\usepackage{booktabs}
\usepackage[separate-uncertainty=true]{siunitx}
\usepackage{amssymb}
\usepackage[figuresright]{rotating}
\usepackage{mathtools}

\newcommand{\dd}{\mathrm{d}}
\newcommand{\abs}[1]{\lvert#1\rvert}

\newcolumntype{d}[1]{D{.}{.}{#1} }
\DeclareMathOperator{\Real}{Re}
\DeclareMathOperator{\Imaginary}{Im}
\DeclareMathOperator{\diag}{diag}
\DeclareSIUnit\parsec{pc}

\begin{document}

\title{Planck-scale constraints on anisotropic Lorentz and CPT invariance violations from optical polarization measurements}
\date{\today}

\author{Fabian Kislat}
\email[Send correspondence to: ]{fkislat@physics.wustl.edu}
\author{Henric Krawczynski}
\affiliation{Washington University in St.\ Louis, Department of Physics and McDonnell Center for the Space Sciences, St.\ Louis, MO 63130}



\begin{abstract}
  Lorentz invariance is the fundamental symmetry of Einstein's theory of special relativity, and has been tested to great level of detail.
  However, theories of quantum gravity at the Planck scale indicate that Lorentz symmetry may be broken at that scale motivating further tests.
  While the Planck energy is currently unreachable by experiment, tiny residual effects at attainable energies can become measurable when photons propagate over sufficiently large distances.
  The Standard-Model Extension (SME) is an effective field theory approach to describe low-energy effects of quantum gravity theories.
  Lorentz and CPT symmetry violating effects are introduced by adding additional terms to the Standard Model Lagrangian.
  These terms can be ordered by the mass dimension of the corresponding operator, and the leading terms of interest have dimension $d=5$.
  Effects of these operators are a linear variation of the speed of light with photon energy, and a rotation of the linear polarization of photons quadratic in photon energy, as well as anisotropy.
  We analyzed optical polarization data from 72 AGN and GRBs and derived the first set of limits on all 16 coefficients of mass dimension $d=5$ of the SME photon sector.
  Our constraints imply a lower limit on the energy scale of quantum gravity of $10^6$ times the Planck energy, severly limiting the phase space for any theory that predicts a rotation of the photon polarization quadratic in energy.
\end{abstract}

\maketitle

\section{Introduction}
Lorentz invariance, the fundamental symmetry of Einstein's theory of special relativity, has been established by many classic experiments, such as the Michelson-Morley experiment~\cite{aa_michelson_ew_morley_1887}, and since then been tested to a great level of detail~\cite{[][{ (2016).}]va_kostelecky_n_russell_2016}. 
However, theories that attempt to unify gravity and the Standard Model of particle physics at the Planck scale~($E_P = \sqrt{c^5\hbar/G} \approx \SI{1.22e19}{GeV}$), imply that there may be deviations from Lorentz invariance at these energies~\cite{g_amelino_camelia_etal_1998,*lj_garay_1998,*r_gambini_j_pullin_1999,*d_mattingly_2005,*t_jacobson_etal_2006,*s_liberati_l_maccione_2009}.
These predictions motivate even more detailed tests of Lorentz invariance.

In the photon sector, violations of Lorentz invariance can lead to an energy dependent vacuum dispersion and birefringence, as well as an anisotropy of the vacuum~\cite{va_kostelecky_m_mewes_2008}.
At attainable energies, $E \ll E_P$, any deviation from Lorentz symmetry is expected to be very small.
However, when photons travel over cosmological distances, even tiny deviations accumulate resulting in potentially measurable effects~\cite{g_amelino_camelia_etal_1998,*lj_garay_1998,*r_gambini_j_pullin_1999,*d_mattingly_2005,*t_jacobson_etal_2006,*s_liberati_l_maccione_2009}.

Vacuum birefringence leads to a wavelength-dependent rotation of the polarization vector of linearly polarized photons.
Hence, broadband polarimetric observations of astrophysical sources can be used to test Lorentz invariance.
Such measurements are generally more sensitive than dispersion measurements by the ratio between the period of the light wave and the resolution with which the arrival times can be measured in a dispersion test, see e.\,g.~\cite{va_kostelecky_m_mewes_2009}.
Note that the latter is usually limited by the source dependent flux variability time scale which can exceed the timing resolution of the detector by several orders of magnitude.

The Standard-Model Extension (SME) is an effective field theory approach describing low-energy effects of a more fundamental theory of physics at the Planck scale, such as Lorentz and CPT violation~\cite{d_colladay_va_kostelecky_1997,*d_colladay_va_kostelecky_1998,*va_kostelecky_2004,va_kostelecky_m_mewes_2009}.
It considers additional terms to the Standard Model Lagrangian, which can be ordered by the mass dimension $d$ of the corresponding tensor operator.
Photon dispersion introduced by operators of dimension $d$ is proportional to $E^{d-4}$, and birefringence is proportional to $E^{d-3}$.
All $(d-1)^2$ coefficients of odd $d$ lead to both dispersion and birefringence, whereas for even $d$ there is a subset of $(d-1)^2$ non-birefringent coefficients.
In a previous paper, we provided the first complete set of constraints on all non-birefringent coefficients of $d=6$ using Fermi observations of AGN light curves~\cite{kislat_f_krawczynski_h_2015}.
In this paper, we use optical polarization measurements to fully constrain the coefficients of $d=5$ setting lower limits on the energy where birefringence due to Lorentz-invariance violation becomes effective well beyond $E_P$.

Very tight constraints on the isotropic Lorentz-invariance violating (LIV) parameters of $d=5$, as well as linear combinations of the anisotropic parameters, have been derived from x-ray polarization measurements of GRBs~\cite{va_kostelecky_m_mewes_2013, k_toma_etal_2012,*laurent_etal_2011,*stecker_2011}.
These limits, however, suffer from a relatively low statistical confidence of the polarization measurements and large systematic uncertainties.
Studies of Cosmic Microwave Background (CMB) polarization data provide strong constraints that do not have these issues and are sensitive to some of the anisotropic parameters~\cite{gubitosi_etal_2009,*va_kostelecky_m_mewes_2007}.

As mentioned above, dimension 5 terms also lead to photon dispersion.
The tightest constraint on linear variations of the speed of light has been derived from time-of-flight measurements of GRB 090510 with the Fermi satellite~\cite{aa_abdo_etal_2009}, placing a lower limit on the relevant energy scale of quantum gravity at $1.22E_P$.
This limit is not competitive in the SME framework.
However, there are certain theories, such as Doubly-Special Relativity (DSR,~\cite{[][{ and references therein.}]g_amelino_camelia_2010}, however, see Sec. IV.F.3 of Ref.~\cite{va_kostelecky_m_mewes_2009} for a critique of DSR), that cannot be described in the effective field theory framework.

Optical polarization measurements are highly sensitive and exist for a sufficient number of sources in the sky to allow a spherical decomposition and, hence, to individually constrain all $d=5$ parameters of the SME photon sector.
We have studied optical polarization from 72 AGN and GRB afterglows and found constraints on each of the 16 parameters of $d=5$ that are stronger than those derived from CMB polarization.

This paper is structured as follows: Section~\ref{sec:framework} gives an overview of the mathematical framework of vacuum birefringence due to Lorentz-invariance violation in the Standard-Model Extension.
Section~\ref{sec:methods} details the methods used in this analysis, which build on this framework.
Section~\ref{sec:results} lists constraints on vacuum birefringence obtained in the present analysis, and we derive constraints on the Lorentz-invariance violating parameters of the Standard-Model Extension.
Section~\ref{sec:summary} summarizes our findings.

\section{Vacuum birefringence in the Standard-Model Extension}\label{sec:framework}
In the Standard-Model Extension the vacuum photon dispersion relation can be written as~\cite{va_kostelecky_m_mewes_2009}
\begin{equation}
  E(p) \simeq \left(1 - \varsigma^0 \pm \sqrt{\bigl(\varsigma^1\bigr)^2 + \bigl(\varsigma^2\bigr)^2 + \bigl(\varsigma^3\bigr)^2}\right) \, p.
\end{equation}
An expansion in mass-dimension and spherical harmonics yields for photons of momentum $p$ arriving from direction $(\theta_k, \varphi_k)$:
\begin{align}\label{eq:sigma0}
  \varsigma^0 &= \sum_{djm}p^{d-4} Y_{jm}(\theta_k, \varphi_k)c_{(I)jm}^{(d)}, \\
  \notag\varsigma^\pm &= \varsigma^1 \pm \varsigma^2 \\
      &= \sum_{djm}p^{d-4} \prescript{}{\mp2}{Y}_{jm}(\theta_k, \varphi_k) \left(k_{(E)jm}^{(d)} \mp ik_{(B)jm}^{(d)}\right), \\
  \varsigma^3 &= \sum_{djm}p^{d-4} Y_{jm}(\theta_k, \varphi_k)k_{(V)jm}^{(d)},
\end{align}
where $k_{(V)jm}^{(d)}$ represents sets of $(d-1)^2$ CPT-odd coefficients, which are non-zero only for odd $d$.
The other coefficients are CPT-even and non-zero only for even $d$.
Hence, in the lowest-order non-minimal SME with $d=5$ there are 16 complex coefficients $k_{(V)jm}^{(5)}$ describing Lorentz invariance violation in the photon sector.
Since $\varsigma^3$ must be real, 
\begin{equation}\label{eq:kVjnegm}
  k_{(V)j-m}^{(d)} = (-1)^m\bigl(k_{(V)jm}^{(d)}\bigr)^*,
\end{equation}
leading to a total of 16 real parameters.

Non-zero values of these parameters will result in an energy, direction and polarization dependence of the photon velocity in the vacuum.
The latter leads to a birefringence of the vacuum.
The polarization angles of two photons observed at energies $E_1$ and $E_2$ emitted at red shift $z_k$ which initially have the same polarization angle will differ at present by~\cite{va_kostelecky_m_mewes_2013}
\begin{equation}\label{eq:delta_psi}
  \begin{split}
    \Delta\psi &= 
        (E_1^2 - E_2^2) \, L_{z_k}^{(5)} \sum_{\substack{j = 0 \ldots 3    \\
                                                         m = -j \ldots j}}
        Y_{jm}(\theta_k,\varphi_k)k_{(V)jm}^{(5)} \\
    &\equiv (E_1^2 - E_2^2) \, \zeta_k^{(5)},
  \end{split}
\end{equation}
where $L_{z_k}^{(5)} = \int_0^{z_k} (1+z)/H_z \, \dd z$ and
\begin{equation}
  H_z = H_0[\Omega_r(1+z)^4 + \Omega_m(1+z)^3 + \Omega_k(1+z)^2 + \Omega_\Lambda]^{\frac{1}{2}}
\end{equation}
with the present day Hubble constant $H_0 = \SI{67.11}{\kilo\meter\per\second\per\mega\parsec} = \SI{1.43e-42}{GeV}$, $\Omega_r = 0.015$, $\Omega_m = 0.317$, $\Omega_\Lambda = 0.686$, and $\Omega_k = 1 - \Omega_r - \Omega_m - \Omega_\Lambda$.
We introduce the parameter $\zeta_k^{(5)}$, which is constrained by the observations.
Hence, a measurement of $\Delta\psi$ allows to contrain a linear combination of the $k_{(V)jm}^{(5)}$.

Spectropolarimetric measurements allow direct application of Eq.~\eqref{eq:delta_psi}.
The main difficulty is then to constrain the rotation angle caused by Lorentz-invariance violation, $\Delta\psi_\text{LIV}$, in the presence of a possible source intrinsic rotation $\Delta\psi_\text{source}$.
This procedure is described in Section~\ref{sub:spectropolarimetry}.

When integrating over an energy range $E_1 \ldots E_2$, on the other hand, one makes use of the fact that a large polarization swing over this energy range would essentially cancel out any observable polarization.
The analysis of spectrally integrated polarization measurements is described in detail in Section~\ref{sub:integratedpolarimetry}.

\section{Methods}\label{sec:methods}
\subsection{Spectropolarimetric measurements}\label{sub:spectropolarimetry}
In our analysis we made use of a large sample of publicly available spectropolarimetric measurements of Active Galactic Nuclei (AGN)~\cite{smith_etal_2009} covering observer frame wavelengths between \SIrange[range-phrase=\ and\ ]{4000}{7550}{\angstrom}.
From this sample we selected distant sources with redshift $z > 0.6$.
We fit the polarization angle of each measurement that resulted in a polarization fraction $P > \SI{10}{\percent}$ with a linear function,
\begin{equation}\label{eq:rotation-rho}
  \psi(\lambda) = \rho_k \, \lambda + C,
\end{equation}
in the wavelength interval $\SI{4500}{\angstrom} \leq \lambda < \SI{7000}{\angstrom}$ in order characterize the change of the polarization angle within this range.
We use a linear fit, instead of a quadratic function as one would expect from Eq.~\eqref{eq:delta_psi}, since the small bandwidth is not sufficiently sensitive to any curvature in the parametrization.

\begin{figure}
  \centering
  \includegraphics[width=.85\columnwidth]{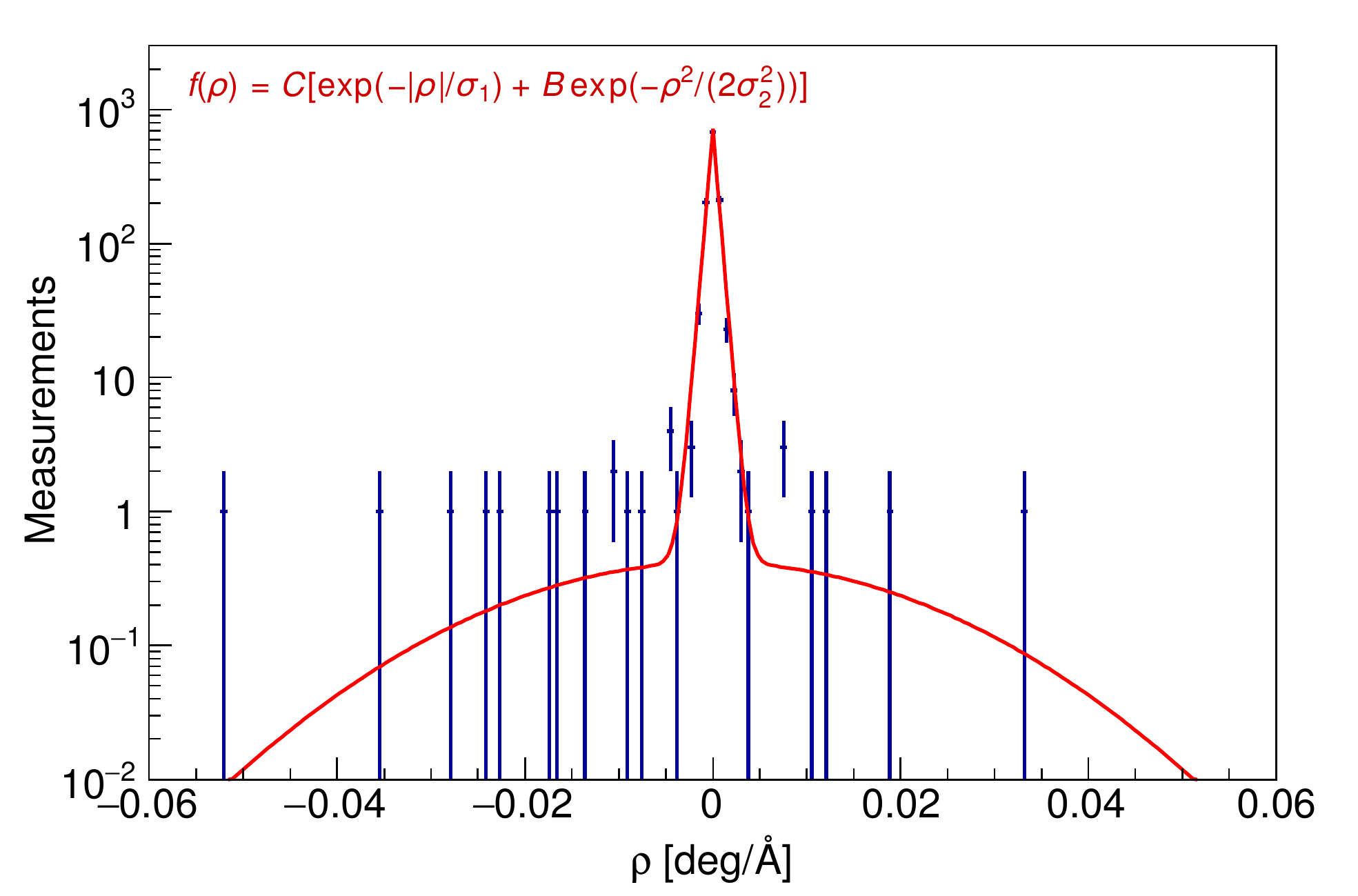}
  \caption{Rotation of the polarization vector in the wavelength range \SIrange{450}{700}{nm} according to Eq.~\eqref{eq:rotation-rho} for AGN with $z < 0.4$ in the Steward Observatory sample~\cite{smith_etal_2009}. The fit parameters are: $C = \num{720(25)}$, $B = \num{0.58(16)e-4}$, $\sigma_1 = \SI{5.17(15)e-4}{deg/\angstrom}$, $\sigma_2 = \SI{1.88(29)e-2}{deg/\angstrom}$. We used a binned log-likelihood fit that correctly takes into account empty bins. As a measure of the goodness of fit, we calculated the chi-squared using a weight of $1$ for empty bins, resulting in $\chi^2/N_{\mathrm{df}} = 73/155$.}
  \label{fig:rotation-nearby}
\end{figure}

In order to determine whether there is a significant influence on~$\psi$ due to Lorentz invariance violation, we compare the distribution of fit parameters $\rho_k$ from each source with values obtained from ``nearby'' sources.
As a control sample, we use the distribution of $\rho$ values from all polarization measurements with $P > \SI{10}{\percent}$ of sources with $z < 0.4$.
This sample consists of the following 16 objects: 3C 66A, BL Lac, MG1 J021114+1051, Mrk 421, OJ 287, ON 325, PKS 0736+01, PKS 1222+216, PKS 1510--08, PKS 2155--304, PKS 2233--148, PMN J0017--0512, S3 1227+25, S4 0954+658, S5 0716+714, and W Comae.
The resulting distribution and its parametrization, $f(\rho)$, are shown in Fig.~\ref{fig:rotation-nearby}.

It should be noted that the source classes of distant and nearby objects are not identical.
Almost all distant sources are flat-spectrum radio quasars (FSRQs), with the only exception being the BL Lac SDSS J084411+5312.
The majority of nearby AGN, on the other hand, are BL Lac-type sources.
This is due to a well-known observational bias: BL Lacs are more abundant but generally weaker than FSRQs~\cite{ackermann_etal_2015}.
Reproducing Fig.~\ref{fig:rotation-nearby} for the two source classes individually resulted in no significant difference in the parameter $\sigma_1$ of $f(\rho)$.
There are not enough data points in the individual distributions to constrain the parameter $\sigma_2$.
We, therefore, argue that this difference in source population will not have a significant influence on the search for Lorentz invariance violation.

We use the parametrization $f(\rho)$ to define the logarithm of the likelihood ratio,
\begin{equation}
  \Lambda_k(\rho_L) = \mathcal{L}_k(\rho_L) - \mathcal{L}_k(0)
\end{equation}
with
\begin{equation}
  \mathcal{L}_k(\rho) = \sum_{i=1}^{N_k} \ln(f(\rho_{k,i} - \rho)),
\end{equation}
where the sum runs over all $N_k$ polarization measurements $\rho_{k,i}$ from the $k$th source.
The LIV-induced rotation of the polarization angle is given by
\begin{equation}\label{eq:rho_L}
  \rho_L = \frac{\Delta(E^2)}{\Delta\lambda}\zeta_k^{(5)}
\end{equation}
with
\begin{equation}
  \Delta(E^2) = \left(\frac{hc}{\lambda_2}\right)^2 - \left(\frac{hc}{\lambda_1}\right)^2
\end{equation}
and the bandwidth $\Delta\lambda = \SI{2500}{\angstrom}$.
We reject a given value $\rho_L$ at the \SI{95}{\percent} confidence level if $\Lambda_k \leq \eta_L$ for $\eta_L$ chosen such that the probability $P(\Lambda_k \leq \eta_L | \rho_L) = 0.05$ under the assumption that there is an LIV-induced rotation $\rho_L$.
This probability, and hence the value of $\eta_L$ that must be chosen, depends on the value of $\rho_L$ and the number of measurements $N_k$.
We find this value using Monte Carlo integration by generating $10^5$ data sets each consisting of $N_k$ measurements of $\rho$ randomly drawn from the distribution $f(\rho - \rho_L)$ for a given value $\rho_L$.

Upper and lower limits $\rho^{(k)}_{L,\text{min/max}}$ are then determined from the observations of the $k$th source by finding the smallest value $\rho^{(k)}_{L,\text{max}} > 0$ and the largest value $\rho^{(k)}_{L,\text{min}} < 0$, for which $\Lambda_k(\rho^{(k)}_L) \leq \eta_L$.
These constraints can directly be converted into limits on the LIV parameter $\zeta_k^{(5)}$ using Eq.~\eqref{eq:rho_L}.

Note that our method is conservative in the sense that Lorentz invariance violations in the nearby source sample lead to a broader $f(\rho)$ distribution, and thus lead to weaker upper limits.

\subsection{Spectrally integrated polarization measurements}\label{sub:integratedpolarimetry}
When integrating over an energy range $E_1 \ldots E_2$ in a polarization measurement, the observed polarization will essentially vanish if $\Delta\psi > \pi$ across the observed energy range, independent of the polarization fraction at the source.
This has been used in Ref.~\cite{va_kostelecky_m_mewes_2013} to derive constraints on the linear combinations of $k_{(V)jm}^{(5)}$ from gamma-ray polarization measurements of GRBs.
The problem with this approach is that the observed polarizations may not cancel entirely due to an energy-dependence of the photon detection efficiency and the photon spectrum of the source, leading to a residual polarization fraction.

\begin{figure}
  \centering
  \includegraphics[width=\columnwidth]{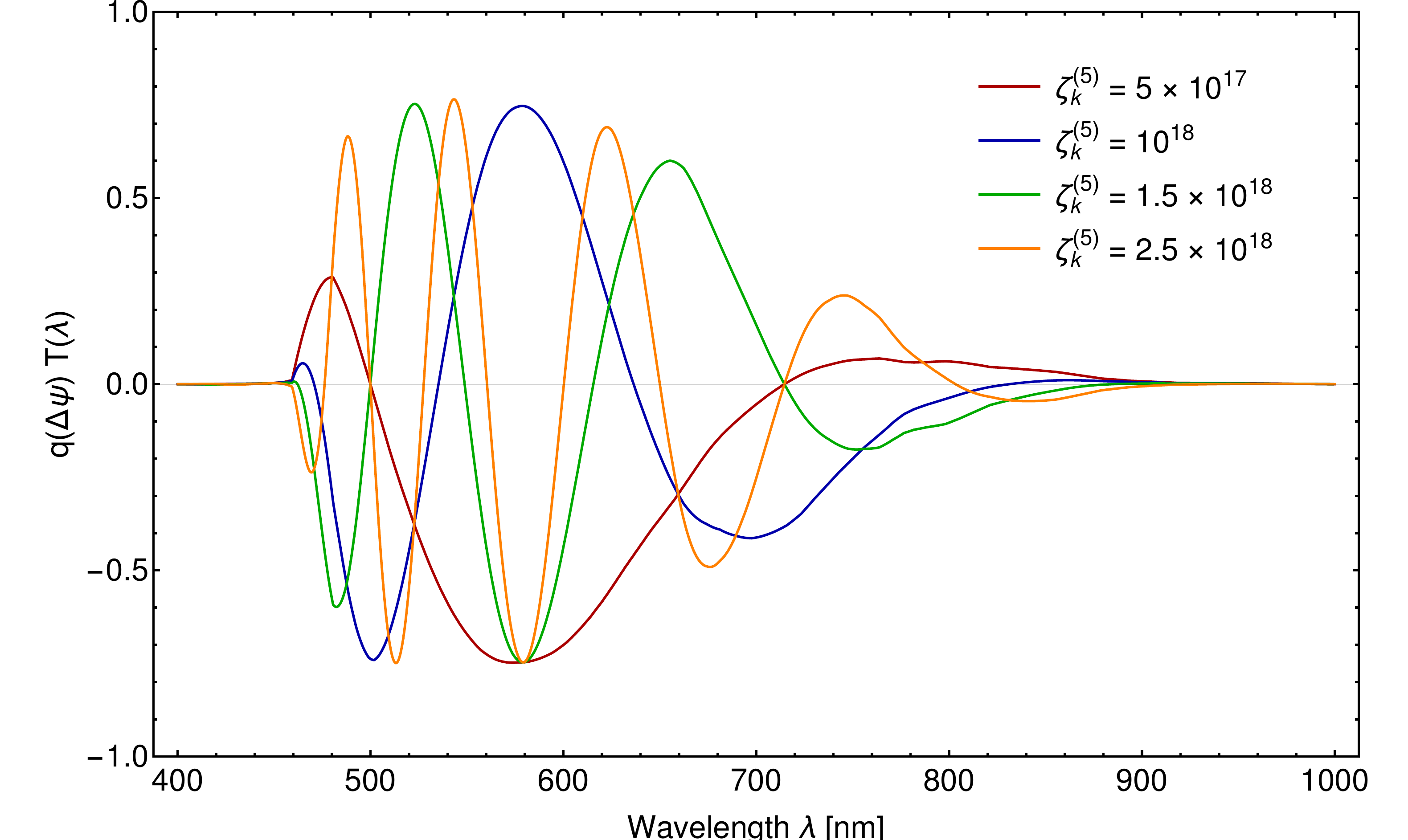}
  \caption{Change of the Stokes parameter $q = \cos(2 \Delta\psi)$ according to Eq.~\eqref{eq:delta_psi} as a function of wavelength folded with the transmissivity of the RINGO ``V+R'' filter~\cite{RINGO,*RINGO2} for a few representative values of $\zeta_k^{(5)}$.}
  \label{fig:q_RINGO}
\end{figure}

Assuming a non-zero parameter $\zeta_k^{(5)}$, an upper limit on the observable polarization in the presence of Lorentz invariance violation can be calculated by integrating the induced rotation of the polarization angle, Eq.~\eqref{eq:delta_psi}, over the bandwidth of the filter.
Assuming a $100\%$ polarized source leads to the following values of the Stokes parameters describing linear polarization:
\begin{align}
  I &= \int\limits_{E_1}^{E_2} T(E) \, \dd E, \\
  Q(\zeta_k^{(5)}) &= \int\limits_{E_1}^{E_2} \cos\bigl(2 \zeta_k^{(5)} (E^2 - E_1^2)\bigr) \, T(E) \, \dd E, \\
  U(\zeta_k^{(5)}) &= \int\limits_{E_1}^{E_2} \sin\bigl(2 \zeta_k^{(5)} (E^2 - E_1^2)\bigr) \, T(E) \, \dd E,
\end{align}
where $T(E)$ is the transmissivity of the filter used for the observation as a function of photon energy, $E = hc/\lambda$.
The integrand of $Q(\zeta_k^{(5)})$ is illustrated in Fig.~\ref{fig:q_RINGO} for the RINGO ``V+R'' filter~\cite{RINGO,*RINGO2}.
In principle, one has to consider the photon spectrum $F(E)$ in addition to the transmissivity.
However, due to the relatively flat optical spectra and narrow bandwidths considered here, this will have only a minor effect.
The source spectra were, thus, neglected in order to simplify the analysis.

\begin{figure}
  \centering
  \includegraphics[width=\columnwidth]{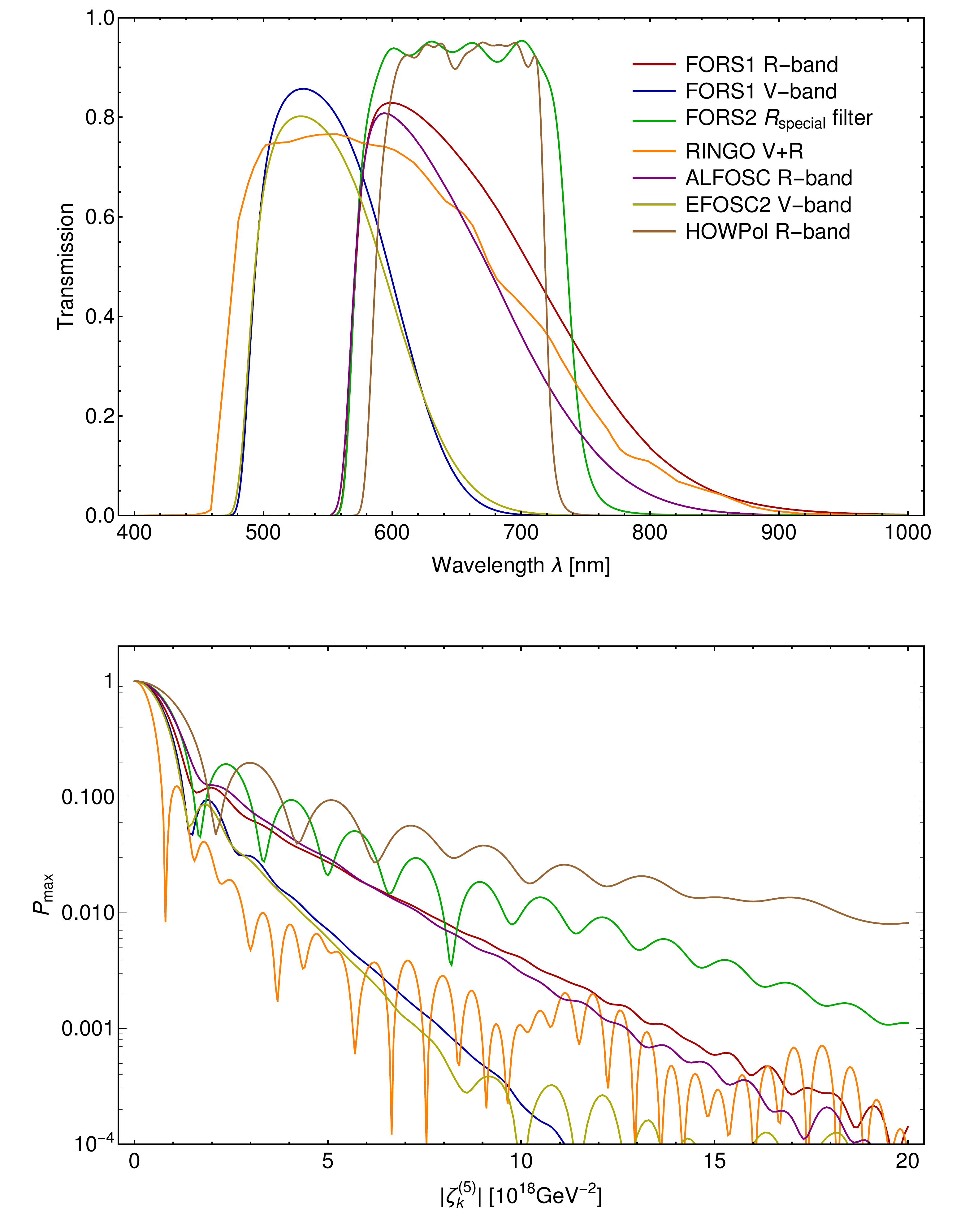}
  \caption{\emph{Top:} Transmissivity of the filters used for the polarization measurements which were used for the analysis in this work~\cite{RINGO,*RINGO2,ALFOSCR,FORS,FORS2,EFOSC2,HOWPol} (note that RINGO and RINGO2 use the same filter). \emph{Bottom:} Upper limit on the observable net polarization according to Eq.~\eqref{eq:pmax} when using these filters as a function of the parameter $\zeta_k^{(5)}$ defined in Eq.~\eqref{eq:delta_psi}.}
  \label{fig:transmission-zeta-5}
\end{figure}

From these equations follows the upper limit on the observable polarization:
\begin{equation}\label{eq:pmax}
  P_\text{max}(\zeta_k^{(5)}) = \frac{\sqrt{Q^2(\zeta_k^{(5)}) + U^2(\zeta_k^{(5)})}}{I}.
\end{equation}
The filter transmission curves used in the measurements which comprised the analysis presented here and resulting maximum observable polarization values, $P_\text{max}$ are shown in  Fig.~\ref{fig:transmission-zeta-5}.
We chose the integration range $E_1 = \SI{1.2}{eV}$ and $E_2 = \SI{2.8}{eV}$, broad enough for all filters.
Filters with a broader bandwidth clearly result in a smaller net polarization.
The relatively flat-top $\mathrm{R}_\text{special}$, V+R, and HOWPol V-band filters result in fringes in the $P_\text{max}$ curves with minima where $\Delta\psi(\zeta_k^{(5)})$ is a multiple of $\pi$ over the bandwidth of the filter.

A measurement of the polarization fraction $P_k$ from the $k$th source can then be converted into a limit on $\zeta_k^{(5)}$ by finding the largest value of $|\zeta_{k,\text{max}}^{(5)}|$, which allows a polarization fraction $P_\text{max} > P_k - 2\sigma_k$, where $\sigma_k$ is the uncertainty of the polarization measurement.

\subsection{Combining multiple measurements to constrain anisotropic Lorentz invariance violation}\label{sub:combiningmeasurements}
The limits on $\zeta_k$ found using the methods described in the previous sections can be converted into constraints on linear combinations of the SME parameters $k_{(V)jm}^{(5)}$ using the definition of $\zeta_k^{(5)}$:
\begin{equation}\label{eq:gamma_k}
  \gamma_{k,\text{max}} = \frac{\zeta_{k,\text{max}}^{(5)}}{L_{z_k}^{(5)}}.
\end{equation}

Since there are 16 real components comprising the SME parameters $k_{(V)jm}^{(5)}$, we use polarization measurements from $N \geq 16$ photon arrival directions, i.\,e. sources in the sky.
This results in a linear system of inequalities derived from Eq.~\eqref{eq:delta_psi}:
\begin{equation}\label{eq:gamma_max}
  \Bigl|\sum_{jm}Y_{jm}(\theta_k, \varphi_k)k_{(V)jm}^{(5)}\Bigr| < \gamma_{k,\text{max}}.
\end{equation}
Constraints on the independent components $v_i^{(5)}$ of $k_{(V)jm}^{(5)}$ can be found using a linear least-squares fit, where $\theta_k$ and $\varphi_k$ are the independent variables, $\gamma_k = 0 \pm \sigma_k$ with $\sigma_k=\gamma_{k,\text{max}}/1.96$ are the measurements, and $v_i^{(5)}$ are the parameters of the fit.
The covariance matrix of these parameters is then given by~\cite{pdg}:
\begin{equation}\label{eq:covariance_matrix}
  \mathbf{V}(v) = (\mathbf{H}^T\mathbf{V}(\gamma)\mathbf{H})^{-1},
\end{equation}
where $\mathbf{H}$ is the coefficient matrix derived from Eq.~\eqref{eq:gamma_max} and $\mathbf{V}(\gamma) = \diag(\sigma_k^2)$ is the covariance matrix of the measurements.
Limits on the components $v_i^{(5)}$ of $k_{(V)jm}^{(5)}$ at the \SI{95}{\percent} confidence level are then given by the diagonal elements of $\mathbf{V}(v)$: $1.96 \times \sqrt{V_{ii}(v)}$.

\section{Results}\label{sec:results}
\begin{figure}
  \centering
  \includegraphics[width=\columnwidth]{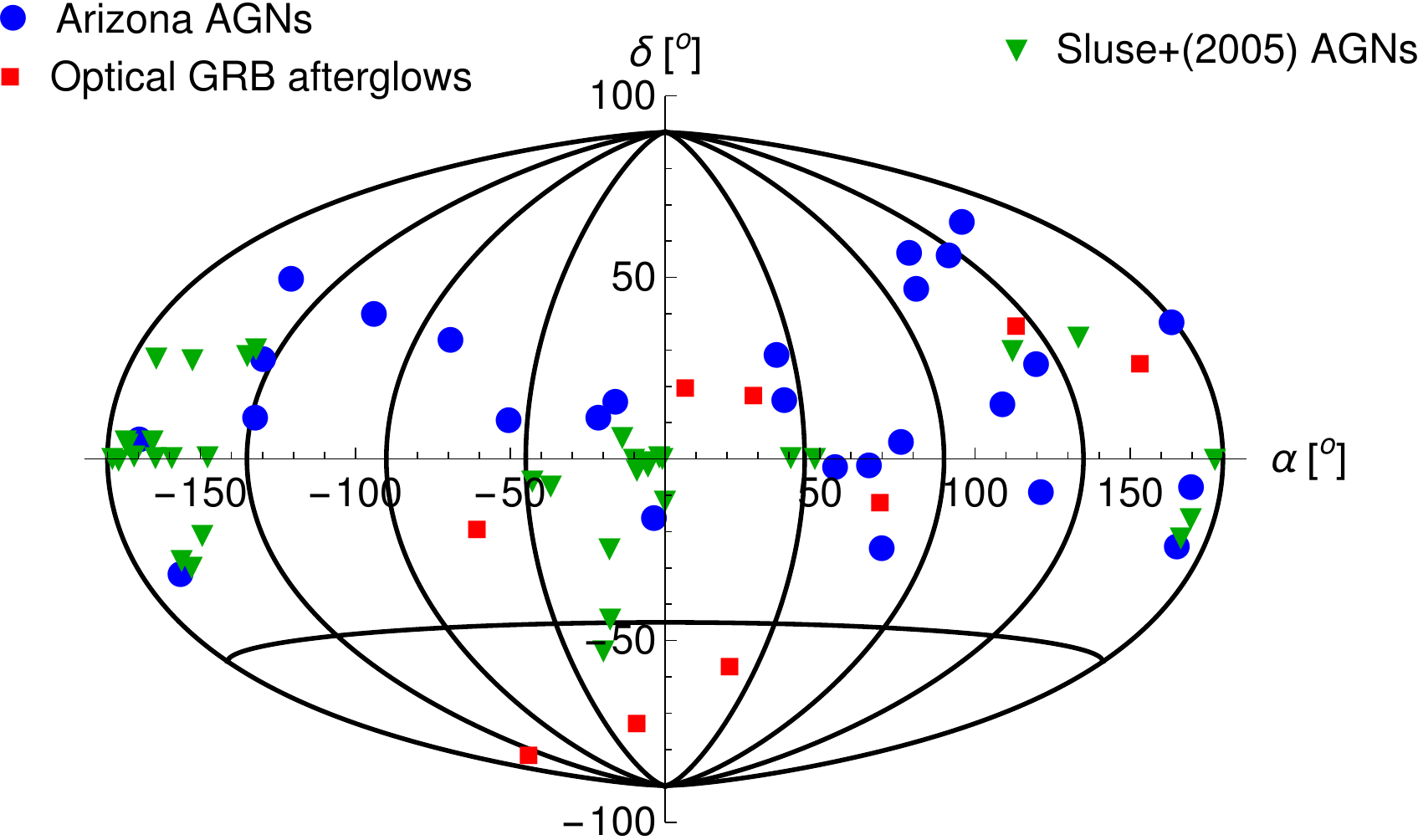}
  \caption{Skymap with all sources used in this analysis. In this paper we derive constraints on Lorentz-invariance violating parameters from optical polarization measurements of GRB afterglows and AGN. The blue circles are AGN selected from the Steward Observatory spectropolarimetric monitoring program~\cite{smith_etal_2009} and the green triangles are sources from the catalog by Sluse et al.~\cite{sluse_etal_2005,*sluse_catalog}. Optical GRB afterglow polarization measurements were collected from various sources~\cite{covino_gotz_2016, wijers_etal_1999,*covino_etal_1999a,*covino_etal_1999b, rol_etal_2000, rol_etal_2003, greiner_etal_2003, gorosabel_etal_2004,*barth_etal_2003, steele_etal_2009,*amati_etal_2009, wiersema_etal_2012, uehara_etal_2012, wiersema_etal_2014}.}
  \label{fig:skymap}
\end{figure}

In this section we provide the constraints on the energy-dependent rotation of the linear polarization direction derived in this work.
Figure~\ref{fig:skymap} shows a skymap of all sources studied in this paper.
It is obvious, that the distribution is not uniform in the sky, but favors small declinations $|\delta| < \SI{45}{\degree}$.
This bias is, amongst others, due to the difficulty of observing extragalactic sources near the galactic plane, which fills more area near the celestial poles than the equator.

We selected all spectropolarimetric observations of sources with redshift $z > 0.6$ during which a polarization fraction $P > \SI{10}{\percent}$ was observed in the Steward Observatory AGN monitoring program.
During cycles 1--7 of the program 27 sources fulfilled these criteria.
Following the procedure described in Section~\ref{sub:spectropolarimetry}, we derived limits on $\zeta_k^{(5)}$ and $\gamma_k$ for all sources.
All sources and corresponding limits are listed in Table~\ref{tab:smith-sources}.

\begin{table*}
  \centering
  \caption{Sources selected from the Steward Observatory spectropolarimetric monitoring project~\cite{smith_etal_2009}. The second column lists the highest observed polarization fraction during cycles 1--7 of the monitoring program. Coordinates have been obtained from the SIMBAD database~\cite{m_wenger_etal_2000}. Individual references are given for the red shifts. The last two columns give the constraints on birefringence due to Lorentz invariance violation derived from these observations.}
  \begin{tabular}{
    l
    S[table-format=2.2]
    S[table-format=3.3]
    S[table-format=+3.3,retain-explicit-plus]
    S[table-format=2.3]
    l
    S[table-format=3.3]
    S[table-format=3.3]
  }
  \toprule
    \multicolumn{1}{c}{\textbf{Source}}               &
    \multicolumn{1}{c}{$\bm{P_\text{max}}$}           &
    \multicolumn{1}{c}{\textbf{RA}}                   &
    \multicolumn{1}{c}{\textbf{Dec.}}                 &
    \multicolumn{2}{c}{\textbf{Redshift}}             &
    \multicolumn{1}{c}{$\bm{\zeta_\text{max}^{(5)}}$} &
    \multicolumn{1}{c}{$\bm{\gamma_\text{max}}$}      \\
                                                      &
    \multicolumn{1}{c}{[\%]}                          &
    \multicolumn{1}{c}{J2000 [${}^\circ$]}            &
    \multicolumn{1}{c}{J2000 [${}^\circ$]}            &
    \multicolumn{2}{c}{$z$}                           &
    \multicolumn{1}{c}{[$10^{15}\si{GeV^{-2}}$]}      &
    \multicolumn{1}{c}{[$10^{-27}\si{GeV^{-1}}$]}     \\
  \midrule
    3C 454.3          & 18.83 & 343.491 & +16.148 & 0.859 & \cite{wa_barkhouse_pb_hall_2001} &   3.321  &   5.051 \\
    4C 14.23          & 20.32 & 111.320 & +14.420 & 1.038 & \cite{ms_shaw_etal_2012}         &   2.160  &   2.720 \\
    4C 28.07          & 30.30 & 39.468  & +28.802 & 1.206 & \cite{ms_shaw_etal_2012}         &   3.341  &   3.633 \\
    AO 0235+164       & 39.79 & 39.662  & +16.616 & 0.940 & \cite{ls_mao_2011}               &   4.225  &   5.871 \\
    B2 1633+382       & 27.26 & 248.815 & +38.135 & 1.813 & \cite{kn_abazajian_etal_2009}    &   4.119  &   3.055 \\
    B2 1846+32A       & 28.88 & 282.092 & +32.317 & 0.800 & \cite{ms_shaw_etal_2012}         &   19.14  &  31.28  \\
    B3 0650+453       & 16.16 & 103.599 & +45.240 & 0.928 & \cite{ms_shaw_etal_2012}         &  200.5   & 282.3   \\
    B3 1343+451       & 10.07 & 206.388 & +44.883 & 2.534 & \cite{ms_shaw_etal_2012}         &    1.480 &   0.819 \\
    BZU J0742+5444    & 21.73 & 115.666 & +54.740 & 0.723 & \cite{jp_halpern_etal_2003}      &    9.999 &  18.11  \\
    CTA 26            & 26.21 & 54.879  & -1.777  & 0.852 & \cite{k_enya_etal_2002}          &    8.101 &  12.42  \\
    CTA 102           & 23.97 & 338.152 & +11.731 & 1.037 & \cite{d_donato_etal_2001}        &   15.55  &  19.60  \\
    MG1 J123931+0443  & 33.61 & 189.886 & +4.718  & 1.760 & \cite{kn_abazajian_etal_2009}    &   14.62  &  11.14  \\
    OJ 248            & 18.09 & 127.717 & +24.183 & 0.941 & \cite{kn_abazajian_etal_2009}    &   14.38  &  19.95  \\
    PKS 0420-014      & 28.67 & 65.816  & -1.343  & 0.916 & \cite{dh_jones_etal_2009}        &    1.875 &   2.673 \\
    PKS 0454-234      & 35.27 & 74.263  & -23.414 & 1.003 & \cite{wa_barkhouse_pb_hall_2001} &    2.732 &  35.58  \\
    PKS 0502+049      & 17.59 & 76.347  & +4.995  & 0.954 & \cite{yy_kovalev_etal_1999}      &  106.9   & 146.3   \\
    PKS 0805-077      & 28.27 & 122.065 & -7.853  & 1.837 & \cite{wa_barkhouse_pb_hall_2001} &   11.35  &   8.319 \\
    PKS 1118-056      & 22.54 & 170.355 & -5.899  & 1.297 & \cite{wa_barkhouse_pb_hall_2001} &  243.0   & 246.3   \\
    PKS 1124-186      & 10.49 & 171.768 & -18.955 & 1.048 & \cite{ls_mao_2011}               &    2.705 &   3.375 \\
    PKS 1244-255      & 13.97 & 191.695 & -25.797 & 0.638 & \cite{wa_barkhouse_pb_hall_2001} &    8.194 &  16.87  \\
    PKS 1441+252      & 37.70 & 220.987 & +25.029 & 0.939 & \cite{ms_shaw_etal_2012}         &    0.622 &   0.865 \\
    PKS 1502+106      & 45.16 & 226.104 & +10.494 & 1.839 & \cite{kn_abazajian_etal_2009}    &    6.946 &   5.086 \\
    PKS 2032+107      & 12.36 & 308.843 & +10.935 & 0.601 & \cite{ms_shaw_etal_2012}         &  409.1   & 895.9   \\
    PMN J2345-1555    & 32.69 & 356.302 & -15.919 & 0.621 & \cite{ms_shaw_etal_2012}         &    3.550 &   7.516 \\
    S4 1030+61        & 37.71 & 158.464 & +60.852 & 1.400 & \cite{kn_abazajian_etal_2009}    &    5.786 &   5.450 \\
    SDSS J084411+5312 & 18.72 & 131.049 & +53.214 & 3.704 & \cite{kn_abazajian_etal_2009}    &   13.46  &   5.474 \\
    Ton 599           & 33.16 & 179.883 & +29.246 & 0.724 & \cite{kn_abazajian_etal_2009}    &    2.007 &   3.628 \\
  \bottomrule
  \end{tabular}
  \label{tab:smith-sources}
\end{table*}

\begin{figure}
  \centering
  \includegraphics[width=\columnwidth]{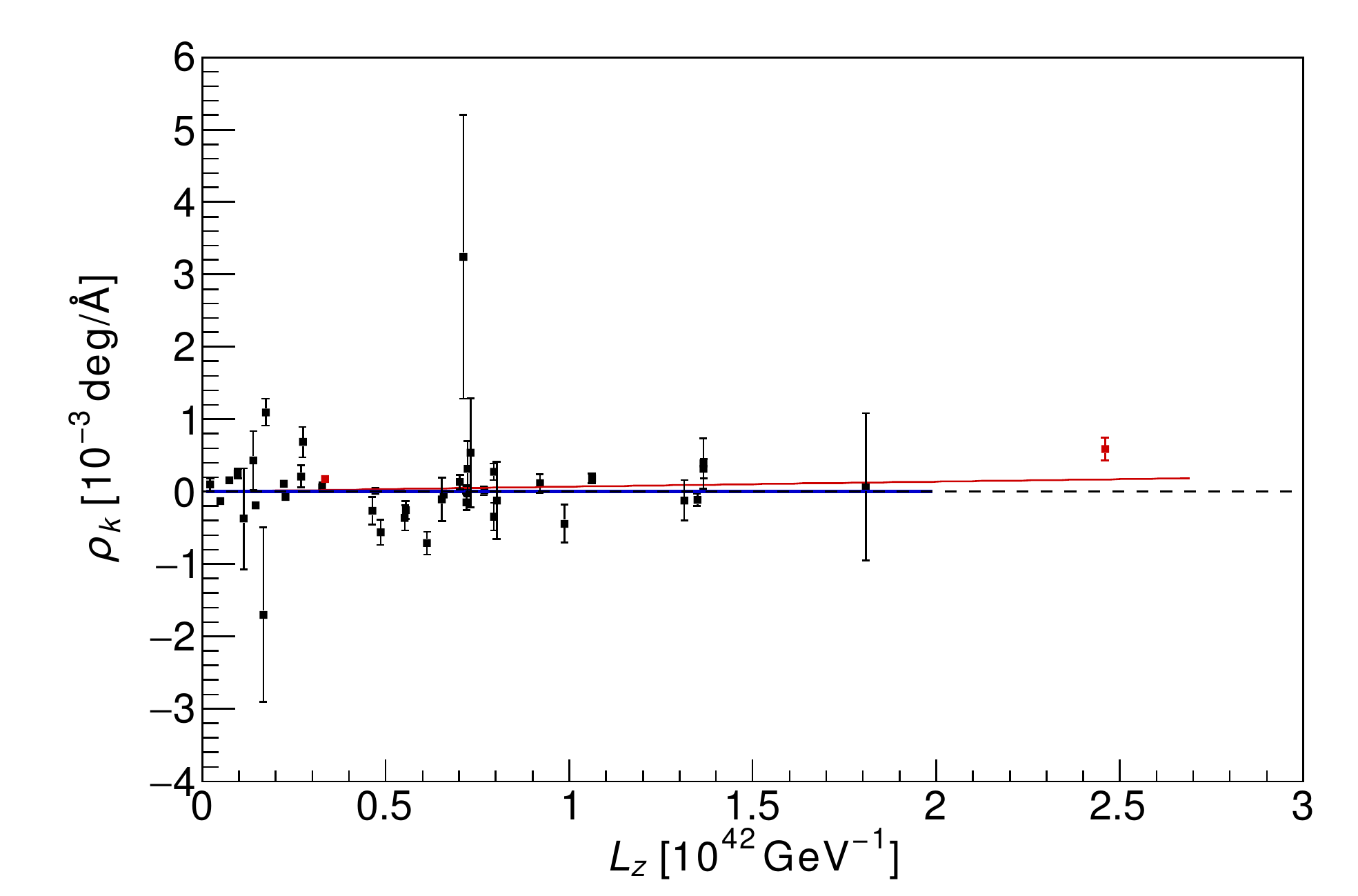}
  \caption{Rotation parameter $\rho$ as a function of the distance measure $L_z$ obtained from all sources in the Steward Observatory AGN monitoring program used in this analysis. Two objects (shown in red) appear to exhibit a source-intrinsic spectral variation of the polarization angle as we argue in the text. They have been removed from the fit, resulting in the thick blue curve.}
  \label{fig:rotation-distance}
\end{figure}

As an additional check, we show the redshift dependence of the average rotation parameter $\rho$ for each source in Fig.~\ref{fig:rotation-distance}.
Each point in this figure corresponds to the weighted average value from all observations of a source.
The error bars have been corrected by incorporating the variance of the measurements to account for variations of $\rho$ for a single source.

We used a simple linear fit to test for the existence of a redshift dependent trend and find a ${\sim}3\sigma$ deviation from zero.
However, the fit quality is rather poor, and we found that this result is entirely dominated by two objects: 3C 66A and SDSS J084411+5312.
Removing the corresponding data points significantly improves the fit quality, while the uncertainty on the slope does not change.
This indicates that these two points do not follow the same linear trend as all other data points.
Since Lorentz-invariance violation is universal, we conclude that these two results must be due to source-intrinsic effects, which justifies their removal from the fit.
We then find a slope of the linear fit of $\rho/L_z = (3.0 \pm 41.5) \times 10^{-50} \, \si{\GeV/\angstrom}$.
This result can be converted into a limit on isotropic Lorentz invariance violation using Eq.~\eqref{eq:delta_psi}: $\abs{k_{(V)00}^{(5)}} < \SI{1.7e-27}{\per\GeV}$ at the \SI{95}{\percent} confidence level and assuming that all other $k_{(V)jm}^{(5)}$ are zero.
This constraint is not competitive because limits derived from X-ray polarization measurements benefit from the significantly larger bandwidth~\cite{va_kostelecky_m_mewes_2013, k_toma_etal_2012,*laurent_etal_2011,*stecker_2011}.

In order to obtain a better coverage of the sky, in particular in the southern hemisphere, we also used spectrally integrated polarization results catalogued in Ref.~\cite{sluse_etal_2005,*sluse_catalog}, as well as optical GRB polarization measurements~\cite{covino_gotz_2016, wijers_etal_1999,*covino_etal_1999a,*covino_etal_1999b, rol_etal_2000, rol_etal_2003, greiner_etal_2003, gorosabel_etal_2004,*barth_etal_2003, steele_etal_2009,*amati_etal_2009, wiersema_etal_2012, uehara_etal_2012, wiersema_etal_2014}.
From the catalog~\cite{sluse_etal_2005,*sluse_catalog} we selected 36 sources for which polarization was measured with at least $5\sigma$ significance.
When choosing GRB afterglow polarization measurements from the literature, we required a $3\sigma$ or greater significance.
Following the procedure described in Section~\ref{sub:integratedpolarimetry}, we use the curve for the appropriate filter in Fig.~\ref{fig:transmission-zeta-5} (bottom) to find the largest value of $\zeta_k^{(5)}$ that may result in a polarization $P_\text{max} \geq P_k - 2\sigma_k$ for each measurement of a polarization fraction $P_k$.
The results and the corresponding limits on $\gamma_k$ according to Eq.~\eqref{eq:gamma_k} are shown in Tables~\ref{tab:sluse-sources} and~\ref{tab:optical-grbs}.

\begin{table*}[p]
  \centering
  \caption{Sources observed by Sluse et al.~\cite{sluse_etal_2005,*sluse_catalog}, including the coordinates and redshifts listed in the reference. All observations were made using the EFOSC2 instrument.}
  \sisetup{separate-uncertainty = false}
  \begin{tabular}{
    l
    S[table-format=3.3]
    S[table-format=+3.3, retain-explicit-plus]
    S[table-format=2.3]
    S[table-format=2.2, table-figures-uncertainty=2]
    S[table-format=3.3]
    S[table-format=3.3]
  }
  \toprule
    \multicolumn{1}{c}{\textbf{Source}}               &
    \multicolumn{1}{c}{\textbf{RA}}                   &
    \multicolumn{1}{c}{\textbf{Dec.}}                 &
    \multicolumn{1}{c}{\textbf{Redshift}}             &
    \multicolumn{1}{c}{\textbf{Polarization}}         &
    \multicolumn{1}{c}{$\bm{\zeta_\text{max}^{(5)}}$} &
    \multicolumn{1}{c}{$\bm{\gamma_\text{max}}$}      \\
                                                      &
    \multicolumn{1}{c}{J2000 [${}^\circ$]}            &
    \multicolumn{1}{c}{J2000 [${}^\circ$]}            &
    \multicolumn{1}{c}{$z$}                           &
    \multicolumn{1}{c}{[\%]}                          &
    \multicolumn{1}{c}{[$10^{18}\si{GeV^{-2}}$]}      &
    \multicolumn{1}{c}{[$10^{-24}\si{GeV^{-1}}$]}     \\
  \midrule
    SDSS J0242+0049   &  40.591 &  +0.820 & 2.071 &  1.47(24) & 4.3 & 2.83 \\
    FIRST03133+0036   &  48.328 &  +0.606 & 1.250 &  1.48(29) & 4.4 & 4.62 \\
    FIRST J0809+2753  & 122.256 & +27.895 & 1.511 &  1.75(20) & 3.9 & 3.42 \\
    PG 0946+301       & 147.421 & +29.922 & 1.220 &  1.65(19) & 4.0 & 4.30 \\
    PKS 1124-186      & 171.768 & -17.045 & 1.048 & 11.68(36) & 1.2 & 1.50 \\
    He 1127-1304      & 172.583 & -12.653 & 0.634 &  1.32(13) & 4.2 & 8.70 \\
    2QZ J114954+0012  & 177.479 &  +0.215 & 1.596 &  1.57(22) & 4.1 & 3.42 \\
    SDSS J1206+0023   & 181.615 &  +0.393 & 2.331 &  0.94(15) & 4.9 & 2.91 \\
    SDSS J1214-0001   & 183.673 &  +0.027 & 1.041 &  2.40(32) & 3.5 & 4.39 \\
    PKS 1219+04       & 185.594 &  +4.221 & 0.965 &  5.56(15) & 2.2 & 2.98 \\
    PKS 1222+037      & 186.218 &  +3.514 & 0.960 &  2.51(22) & 3.3 & 4.49 \\
    TON 1530          & 186.364 & +22.587 & 2.058 &  0.92(14) & 4.9 & 3.25 \\
    SDSS J1234+0057   & 188.616 &  +0.966 & 1.532 &  1.35(23) & 4.4 & 3.81 \\
    PG 1254+047       & 194.250 &  +4.459 & 1.018 &  0.84(15) & 5.1 & 6.55 \\
    PKS 1256-229      & 194.785 & -22.823 & 1.365 & 22.32(15) & 1.0 & 0.97 \\
    SDSS J1302-0037   & 195.534 &  +0.626 & 1.672 &  1.37(20) & 4.3 & 3.43 \\
    PKS 1303-250      & 196.564 & -24.711 & 0.738 &  0.91(17) & 5.0 & 8.87 \\
    FIRST J1312+2319  & 198.056 & +23.333 & 1.508 &  1.10(16) & 4.6 & 4.04 \\
    SDSS J1323-0038   & 200.769 &  +0.649 & 1.827 &  1.13(21) & 4.7 & 3.46 \\
    CTS J13.07        & 205.518 & -17.700 & 2.210 &  0.83(15) & 5.1 & 3.17 \\
    SDSS J1409+0048   & 212.328 &  +0.807 & 1.999 &  3.91(28) & 2.6 & 1.77 \\
    HS 1417+2547      & 215.055 & +25.568 & 2.200 &  1.03(18) & 4.8 & 3.00 \\
    FIRST J1427+2709  & 216.765 & +27.161 & 1.170 &  1.35(25) & 4.5 & 5.04 \\
    FIRST J21079-0620 & 316.990 &  -5.664 & 0.644 &  1.12(22) & 4.8 & 9.79 \\
    SDSS J2131-0700   & 322.912 &  -6.996 & 2.048 &  1.78(32) & 4.1 & 2.73 \\
    PKS 2204-54       & 331.932 & -52.224 & 1.206 &  1.81(26) & 3.9 & 4.24 \\
    PKS 2227-445      & 337.735 & -43.725 & 1.326 &  5.26(48) & 2.4 & 2.38 \\
    PKS 2240-260      & 340.860 & -24.258 & 0.774 & 14.78(21) & 1.1 & 1.86 \\
    PKS 2301+06       & 346.118 &  +6.336 & 1.268 &  3.69(26) & 2.7 & 2.80 \\
    SDSS J2319-0024   & 349.995 &  +0.414 & 1.889 &  1.85(30) & 4.0 & 2.86 \\
    PKS 2320-035      & 350.883 &  -2.715 & 1.411 &  9.56(20) & 1.2 & 1.12 \\
    PKS 2332-017      & 353.835 &  -0.481 & 1.184 &  4.86(19) & 2.3 & 2.55 \\
    PKS 2335-027      & 354.489 &  -1.484 & 1.072 &  3.55(30) & 2.9 & 3.54 \\
    SDSS J2352+0105   & 358.159 &  +1.098 & 2.156 &  1.59(26) & 4.2 & 2.67 \\
    SDSS J2356-0036   & 359.120 &  +0.601 & 2.936 &  1.81(34) & 4.1 & 2.01 \\
    QSO J2359-12      & 359.973 & -11.303 & 0.868 &  4.12(20) & 2.5 & 3.76 \\
  \bottomrule
  \end{tabular}
  \label{tab:sluse-sources}
\end{table*}

\begin{table*}[p]
  \centering
  \caption{Optical GRB measurements.}
  \sisetup{separate-uncertainty = true}
  \begin{tabular}{
    l
    l
    S[table-format=3.3]
    S[table-format=+3.3, retain-explicit-plus]
    S[table-format=2.3]
    S[table-format=2.2, table-figures-uncertainty=2]
    S[table-format=3.3]
    S[table-format=3.3]
    l
  }
  \toprule
    \multicolumn{1}{c}{\textbf{Source}}               &
    \multicolumn{1}{c}{\textbf{Instrument}}           &
    \multicolumn{1}{c}{\textbf{RA}}                   &
    \multicolumn{1}{c}{\textbf{Dec.}}                 &
    \multicolumn{1}{c}{\textbf{Redshift}}             &
    \multicolumn{1}{c}{\textbf{Polarization}}         &
    \multicolumn{1}{c}{$\bm{\zeta_\text{max}^{(5)}}$} &
    \multicolumn{1}{c}{$\bm{\gamma_\text{max}}$}      &
    \multicolumn{1}{c}{\textbf{Refs.}}                \\
                                                      &
                                                      &
    \multicolumn{1}{c}{J2000 [${}^\circ$]}            &
    \multicolumn{1}{c}{J2000 [${}^\circ$]}            &
    \multicolumn{1}{c}{$z$}                           &
    \multicolumn{1}{c}{[\%]}                          &
    \multicolumn{1}{c}{[$10^{18}\si{GeV^{-2}}$]}      &
    \multicolumn{1}{c}{[$10^{-24}\si{GeV^{-1}}$]}     &
                                                      \\
  \midrule
    GRB 990510  & FORS1 R-band & 204.532 & -80.497 & 1.619 & 1.6(2)   & 7.00 &  5.76 & \cite{wijers_etal_1999,*covino_etal_1999a,*covino_etal_1999b} \\
    GRB 990712  & FORS1 R-band & 337.971 & -73.408 & 0.430 & 2.9(4)   & 5.50 & 17.0  & \cite{rol_etal_2000} \\
    GRB 020813  & FORS1 V-band & 296.674 & -19.601 & 1.25  & 1.42(25) & 4.60 & 4.83  & \cite{gorosabel_etal_2004,*barth_etal_2003} \\
    GRB 021004  & NOT/ALFOSC   &   6.728 & +18.928 & 2.330 & 2.1(6)   & 7.50 &  4.46 & \cite{rol_etal_2003} \\
    GRB 030329  & NOT/ALFOSC   & 161.208 & +21.522 & 0.169 & 2.4(4)   & 6.30 & 51.6  & \cite{greiner_etal_2003} \\
    GRB 090102  & RINGO        & 128.248 & +33.107 & 1.547 & 10.2(13) & 1.30 &  1.12 & \cite{steele_etal_2009,*amati_etal_2009} \\
    GRB 091018  & FORS2        &  32.192 & -57.55  & 0.97  & 3.25(35) & 7.55 &  1.02 & \cite{wiersema_etal_2012} \\
    GRB 091208B & HOWPol       &  29.410 &  16.881 & 1.06  & 10.4(25) & 7.30 &  9.00 & \cite{uehara_etal_2012} \\
    GRB 121024A & FORS2        &  70.467 & -12.268 & 2.298 & 4.83(20) & 5.95 &  3.58 & \cite{wiersema_etal_2014} \\
  \bottomrule
  \end{tabular}
  \label{tab:optical-grbs}
\end{table*}

\begin{figure}
  \centering
  \includegraphics[width=.8\columnwidth]{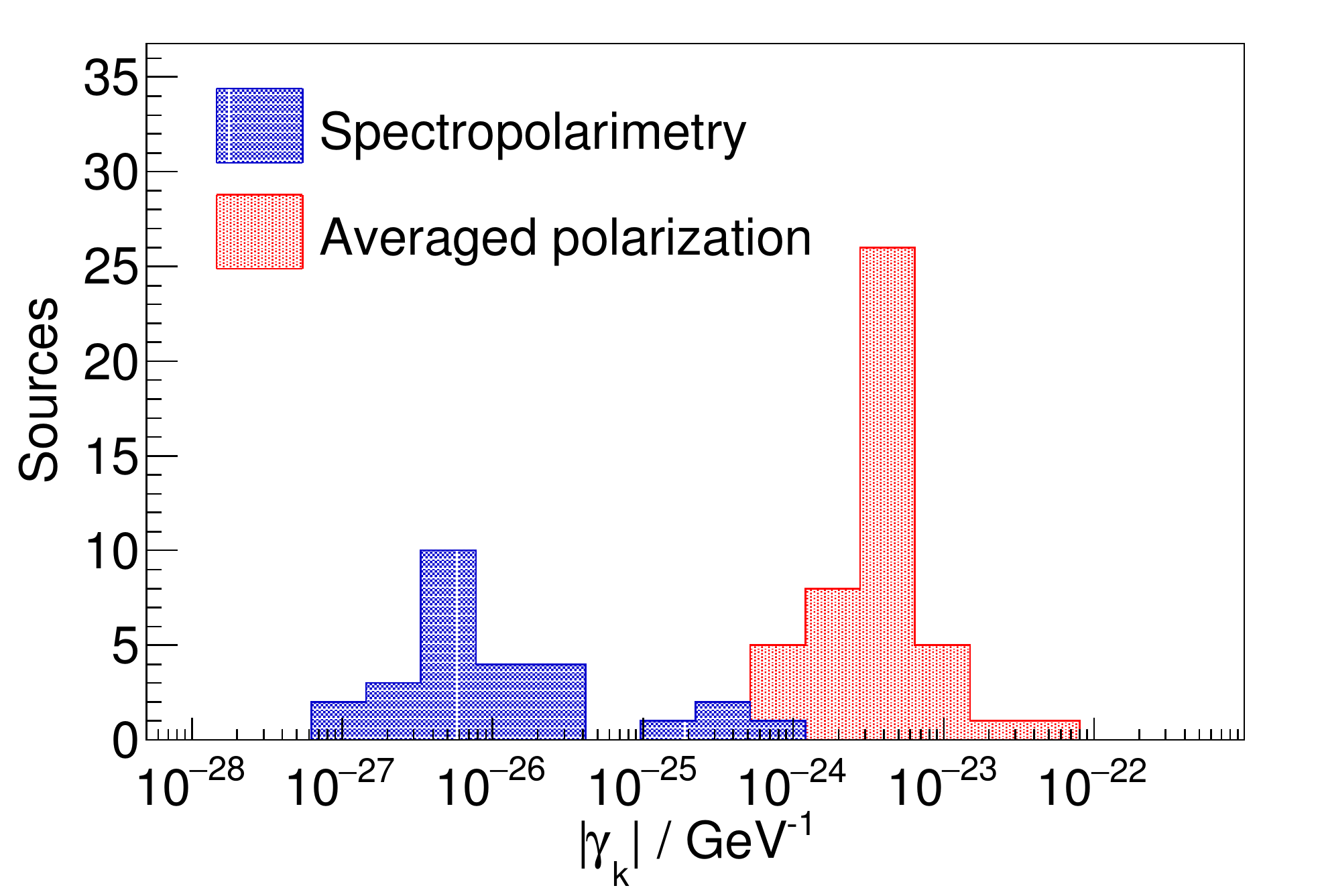}
  \caption{Distribution of all limits $\gamma_k$ from Tables~\ref{tab:smith-sources}--\ref{tab:optical-grbs}. The values can be compared directly to those listed in Ref.~\cite{va_kostelecky_n_russell_2016}. As expected, spectropolarimetric are more sensitive than measurements in which only the spectrally integrated net polarization is considered.}
  \label{fig:sources_gamma}
\end{figure}

Figure~\ref{fig:sources_gamma} shows the distribution of all limits on~$\gamma_k$ obtained in this analysis.
As expected, the spectropolarimetric measurements are most constraining since the rotation of the polarization can be observed directly.
Spectrally integrated polarization measurements, on the other hand, only allow one to deduce an upper limit on the possible rotation.
The values in this table can be compared directly with previously published limits as listed in~\cite{va_kostelecky_n_russell_2016}.
While limits derived from x-ray polarization measurements are about~$7$ orders of magnitude lower than the ones obtained here, the significance of the underlying polarization measurements is much lower than that of the optical polarization measurements used here.
Our limits are more constraining than those derived from CMB polarization measurements, and they are systematically independent.

Using the method described in Section~\ref{sub:combiningmeasurements}, we calculated the covariance matrix of the 16 real components of the~$k_{(V)jm}^{(5)}$, Eq.~\eqref{eq:covariance_matrix}.
The diagonal components of this matrix result in the \SI{95}{\percent} confidence level limits listed in Table~\ref{tab:liv-parameter-limits}.
The constraints on all parameters are equivalent to a lower limit on the energy where Lorentz invariance is violated significantly that exceeds $E_P$ by more than 6 orders of magnitude.

\begin{table}
  \centering
  \caption{Limits at the \SI{95}{\percent} confidence level on all independent SME parameters $k_{(V)jm}^{(5)}$ obtained in this analysis in \si{GeV^{-1}}. The dependent parameters $k_{(V)j-m}^{(5)}$ can be calculated according to Eq.~\eqref{eq:kVjnegm}.}
  \setlength\extrarowheight{1ex}
  \begin{tabular}{
    l@{ $<$ }
    S[table-format=1.1e+4]
    }
  \toprule
    $|k_{(V)00}^{(5)}|$                       & 5.0e-26 \\
    $|k_{(V)10}^{(5)}|$                       & 6.5e-26 \\
    $|\Real\bigl(k_{(V)11}^{(5)}\bigr)|$      & 7.8e-27 \\
    $|\Imaginary\bigl(k_{(V)11}^{(5)}\bigr)|$ & 2.9e-26 \\
    $|k_{(V)20}^{(5)}|$                       & 3.0e-26 \\
    $|\Real\bigl(k_{(V)21}^{(5)}\bigr)|$      & 7.5e-27 \\
    $|\Imaginary\bigl(k_{(V)21}^{(5)}\bigr)|$ & 2.4e-26 \\
    $|\Real\bigl(k_{(V)22}^{(5)}\bigr)|$      & 2.1e-26 \\
    $|\Imaginary\bigl(k_{(V)22}^{(5)}\bigr)|$ & 1.7e-26 \\
    $|k_{(V)30}^{(5)}|$                       & 2.1e-26 \\
    $|\Real\bigl(k_{(V)31}^{(5)}\bigr)|$      & 1.4e-26 \\
    $|\Imaginary\bigl(k_{(V)31}^{(5)}\bigr)|$ & 1.0e-26 \\
    $|\Real\bigl(k_{(V)32}^{(5)}\bigr)|$      & 8.3e-27 \\
    $|\Imaginary\bigl(k_{(V)32}^{(5)}\bigr)|$ & 1.4e-26 \\
    $|\Real\bigl(k_{(V)33}^{(5)}\bigr)|$      & 9.7e-27 \\
    $|\Imaginary\bigl(k_{(V)33}^{(5)}\bigr)|$ & 5.8e-27 \\
  \bottomrule
  \end{tabular}
  \label{tab:liv-parameter-limits}
\end{table}

\section{Summary}\label{sec:summary}
In the Standard-Model Extension (SME), Lorentz invariance violating effects are described by adding additional terms to the Standard Model Lagrangian.
These terms can be ordered by the mass dimension $d$ of the corresponding operator.
Operators of odd dimension are CPT-odd, and operators with even $d$ are CPT-even.
Since terms of $d \leq 4$ are not suppressed by positive powers of $E/E_P$, the leading order of interest for most Lorentz and CPT violation searches is $d = 5$.
Operators of this dimension lead to an energy dependent vacuum dispersion and vacuum birefringence.
In general, these effects can be anisotropic, and a spherical decomposition results in 16 complex coefficients at mass dimension $d=5$ described by 16 real parameters.

We analyzed optical spectropolarimetry data from 27 AGN to derive constraints on vacuum birefringence.
We derived an additional set of constraints from spectrally integrated polarization measurements from 36 AGN and afterglows of 9 GRBs.
Using multiple sources in the sky allows us to perform a spherical decomposition of the LIV constraints and, thus, to constrain the 16 parameters of the SME photon sector at $d=5$.
The results are listed in Table~\ref{tab:liv-parameter-limits}.

All parameters are constrained at the order of a few $10^{-26}\si{GeV^{-1}}$ or better corresponding to a lower limit on the energy scale at which there may be significant Lorentz invariance or CPT violation of $10^6E_P$.
Since the Planck mass is considered an upper limit on the relevant energy scale for quantum gravity in most scenarios, our results severly constrain any theory that predicts an isotropic or anisotropic vacuum birefringence that is quadratic in photon energy.
Furthermore, according to the SME, theories that cause a variation of the speed of light that is linear with photon energy are severly constrained.
This statement holds for all theories whose low-energy effects can be described by an effective field theory.

Tighter constraints than those presented here have been derived from x-ray polarization measurements of GRBs.
However, those measurements suffer from low statistical significance and large systematic uncertainties.
Furthermore, our results for the first time constrain all 16 parameters at $d=5$ individually.
Our results are compatible with those found in CMB polarization measurements, but systematically different.

The methods developed here can be used directly to constrain the birefringent parameters $k_{(E)jm}^{(4)}$ and $k_{(B)jm}^{(4)}$. 
At higher order, however, gamma-ray polarization measurements are necessary.
For example, at $d=6$ it is necessary to measure polarization of $E \simeq \SI{100}{MeV}$ photons to constrain the birefringent parameters at the Planck scale~\cite{kislat_f_krawczynski_h_2015}.
This might be achievable with next-generation Compton and pair-production telescopes (e.\,g.~\cite{sd_hunter_etal_2014, moiseev_etal_2015}).

\section*{Acknowledgements}
We would like to thank Alan Kostelecký, Matthew Mewes, Stefano Covino, Jim Buckley, Floyd Stecker, Paul Smith, and Manel Errando for fruitful discussions.
We are also very thankful to the anonymous referee for the insightful comments and questions.
The authors are grateful for funding from NASA Grant NNX14AD19G and DOE Grant DE-FG02-91ER40628.
Data from the Steward Observatory spectropolarimetric monitoring project were used, which is supported by Fermi Guest Investigator grants NNX08AW56G, NNX09AU10G, NNX12AO93G, and NNX15AU81G.
Furthermore, we made use of the SIMBAD database, operated at CDS, Strasbourg, France.

\bibliography{liv}

\end{document}